\documentclass[aps,floatfix,groupedaddress,nofootinbib,superscriptaddress,twocolumn]{aastex63}
\usepackage[T1]{fontenc}
\usepackage{microtype}
\usepackage{afterpage,amsmath,amssymb,graphicx,natbib, color}
\usepackage{hyperref}
\usepackage[english]{babel}
\usepackage{blindtext}
\usepackage[utf8]{inputenc}
\usepackage{enumitem}
\usepackage{graphicx}
\usepackage{hyperref}
\usepackage{mathtools}
\usepackage{comment}
\usepackage{amsmath}
\usepackage{ amssymb }
\usepackage{lipsum} 
\usepackage{float}
\usepackage{cleveref}
\usepackage[caption=false]{subfig}
\usepackage{placeins}
\usepackage{xcolor}
\usepackage{multirow}
\usepackage{float}

\newcommand{\be}{\begin{equation}}
\newcommand{\ee}{\end{equation}}
\newcommand{\bfig}{\begin{figure}}
\newcommand{\efig}{\end{figure}}

\newcommand{\universemachine}{\texttt{UniverseMachine}}
\newcommand{\limpy}{\texttt{LIMpy}}
\newcommand{\tng}{\texttt{IllustrisTNG}}

\begin{document}
\title{\texttt{$\limpy$}: A Semi-analytic Approach to Simulating Multi-line Intensity Maps at Millimetre Wavelengths}

%% Note that the corresponding author command and emails have to come
%% before everything else. Also, place all the emails in the \email
%% command instead of using multiple \email calls.
\correspondingauthor{Anirban Roy}
\email{ar689@cornell.edu}

\author{Anirban Roy}
\affiliation{Department of Astronomy, Cornell University,
\\ Ithaca, NY 14853, USA}

\author{Dariannette Valentín-Martínez }
\affiliation{Department of Astronomy, University of Florida}
\affiliation{School of Earth and Space Exploration, Arizona State University, Tempe, AZ 85287, USA}

\author{Kailai Wang}
\affiliation{Department of Physics, Cornell University,
\\ Ithaca, NY 14853, USA}

\author{Nicholas Battaglia}
\affiliation{Department of Astronomy, Cornell University,
\\ Ithaca, NY 14853, USA}

\author{Alexander van Engelen}
\affiliation{School of Earth and Space Exploration, Arizona State University, Tempe, AZ 85287, USA}

%-----------------------------------------------------------------------------------%
%Abstract
%-----------------------------------------------------------------------------------%
\begin{abstract}
Mapping of multiple lines such as the fine-structure emission from [CII] (157.7 $\mu \text{m}$), [OIII] (52 \& 88.4 $\mu \text{m}$), and rotational emission lines from CO are of particular interest for upcoming line intensity mapping (LIM) experiments at millimetre wavelengths, due to their brightness features. Several upcoming experiments aim to cover a broad range of scientific goals, from detecting signatures of the epoch of reionization to the physics of star formation and its role in galaxy evolution. In this paper, we develop a semi-analytic approach to modelling line strengths as functions of the star formation rate (SFR) or infrared (IR) luminosity based on observations of local and high-z galaxies.
This package, \texttt{$\limpy$} (Line Intensity Mapping in Python), estimates the intensity
and power spectra of [CII], [OIII], and CO rotational transition lines up to the $J$-levels (1-0) to (13-12) based both on analytic formalism and on simulations. We develop a relation among halo mass, SFR, and
multi-line intensities that permits us to construct a generic formula for the evolution of several line strengths up to $z \sim 10$. We implement a variety of star formation models and multi-line luminosity relations to estimate the astrophysical uncertainties on the intensity power spectrum of these lines. As a demonstration, we predict the signal-to-noise ratio of [CII] detection for an EoR-Spec-like instrument on the Fred Young Submillimeter Telescope (FYST). Furthermore, the ability to use any halo catalogue allows the $\limpy$ code to be easily integrated into existing simulation pipelines, providing a flexible tool to study intensity mapping in the context of complex galaxy formation physics.
\end{abstract}
\keywords{line intensity mapping, galaxy evolution, reionization, structure formation}

\section{Introduction} 
%Emission lines
Observations of redshifted line emissions from atomic and molecular gas in galaxies and intergalactic medium trace the underlying dark matter density fluctuations. Several factors influence the strength of various spectral lines, including the star formation history (SFH), metallicity, and the host halo mass of galaxies. At high redshifts, $z \gtrsim 6$, it is an arduous task to resolve each individual galaxy in a survey field to understand the physics behind galaxy formation, evolution, and their connection to the intergalactic medium. Multi-line intensity mapping (MLIM) encapsulates integrated emissions from both luminous and faint sources, providing rich information about galaxy clustering, star formation rate density (SFRD), and galaxy luminosity functions \citep{Visbal2010, Visbal2011, Kovetz2017LIM_report, Bernal:2022jap}. The detection of different atomic and molecular line intensities at a particular observational frequency probes the Universe at different epochs (or redshifts); thus, it provides a unique opportunity to construct a three-dimensional (3D) map of the Universe by the measurement of several line emission using a couple of observational frequencies.

% Intro to Reionization and connection to LIM \\
Detecting the power spectra of fine structure lines, such as [CII] and [OIII], at high redshift ($z \gtrsim 6$) has the potential to reveal the sources of reionization and their clustering properties \citep{Dumitru2018, Padmanabhan_CII, PadmanabhanAll, Karoumpis2021, Limfast2}. Furthermore, mapping the Universe using various rotational transitions of CO (J-level transition) can probe the formation of structures at high redshifts, offering insights into the process of reionization and the star formation history of the first-generation galaxies \citep{Kovetz2017LIM_report, Breysse:2021ecm}. Employing a tomographic approach to exploring the Universe enables the measurement of key quantities, including the growth factor of structures, the Hubble constant, and the equation of the state of dark energy \citep{Kovetz2017LIM_report, Karkare:2018sar, Bernal:2019gfq, Silva:2019hsh}. A joint analysis of all lines could prove valuable for constraining the inflationary paradigm by limiting $f_{NL}$ \citep{Moradinezhad2018, Bernal:2019jdo, Chen:2021ykb}. Detecting the 21\,cm$-$[CII] cross-power spectrum and cross-bispectrum signals can aid in mitigating the low-redshift contamination of 21\,cm data, and incorporating 21 cm observations may enhance constraints on astrophysical parameters \citep{Beane2018, Dumitru2018, Schaan:2021gzb}.

%Lines in Astrophysical context and advantages \\
Several observational efforts have been made to detect the 21\,cm line emission to study the cosmic dawn, the epoch of reionization (EoR), and late-time structure formation. Intensity mapping of other lines, such as fine-structure emission from carbon [CII] line (157.7 $\mu m$), doubly ionized oxygen [OIII] (88.4 $\mu m$), and rotational emission lines from CO are of particular interest of upcoming LIM experiments \citep{Suginohara1998, Righi2008b, Lidz2011_CO, Carilli2011, Fonseca:2016, Gong2017, Kovetz2017LIM_report, Chung2018CII, PadmanabhanCO, Padmanabhan_CII, Dumitru2018, Chung2018CO, Kannan:2021ucy, Murmu:2021ljb, Karoumpis2021}. Despite the several advantages offered by the MLIM technique, there are key challenges to detecting a particular line emission across a broad redshift range in the presence of foreground and instrumental noise. As many of the lines emitted from other sources can be redshifted to the same observational frequency channel, it will create line confusion by adding extra emission to the particular line emission we aim to detect \citep{Lidz:2016lub, Cheng:2016yvu}. These lines are called \lq interlopers\rq, and their contamination presents an obstacle to the detection of a particular line coming from the sources at a certain redshift. Moreover, the uncertainty of the star formation history (SFH) in galaxies and its relation with the mass of the host halos arises from the lack of observational data, particularly at high redshift when the reionization process occurred. However, the uncertainties in star formation and their relation with the host halos can be well explored by the different high-resolution simulations, such as \textsc{\universemachine} \citep{Behroozi2019}, $\tng$ \citep{TNG-gal}, and \texttt{Emerge} \citep{Moster2018}.

%Current state of experiments (with a focus on CCAT P) 
Multiple experiments like FYST\footnote{\href{https://www.ccatobservatory.org/}{https://www.ccatobservatory.org/}} \citep{CCAT-prime2021}, SPHEREx\footnote{\href{https://spherex.caltech.edu/}{https://spherex.caltech.edu/}} \citep{SPHEREx-science-paper2018}, TIME \citep{Time-science-2014}, CONCERTO\footnote{\href{https://mission.lam.fr/concerto/}{https://mission.lam.fr/concerto/}} \citep{CONCERTO-science-2020}, COMAP\footnote{\href{https://comap.caltech.edu/}{https://comap.caltech.edu/}} \citep{COmap-science-2021}, EXCLAIM \citep{EXCLAIM-2020}, aim to cover a broad range of scientific goals, from the detection of EoR signatures to the formation of stars in galaxies. To explore the synergies among these experiments requires modelling and simulating the desired signal over a broad redshift range. In this work, we develop a package, $\limpy$ \footnote{\href{https://github.com/Anirbancosmo/limpy}{https://github.com/Anirbancosmo/limpy}}, to model and simulate several line emissions up to $z \lesssim 10$. We implement a range of models for the star formation histories and the relations between multi-line luminosity and these star formation histories, based on analytic expressions and simulations. Collecting many models in one place allows us to explore the astrophysical uncertainties of the amplitude and shape of the signals as well as the level of contamination from interlopers. Additionally, we determine the power spectrum of line intensities both analytically through the halo model and from simulations. We adopt a map-making approach by utilizing halo catalogues generated from N-body simulations, which enables us to forecast the detectability of power spectra for future line intensity mapping (LIM) experiments. In our analysis, we incorporate the effects of beam convolution and develop realistic simulations of instrumental noise, comparing the results with those of the analytic halo models. By combining the intensity signals of various lines, $\limpy$ offers valuable insights into the astrophysical processes governing these emissions and their potential detectability in future experiments. Ultimately, this approach paves the way for a deeper understanding of the underlying physics and the potential impact of interlopers on the observed signals.

The $\limpy$ package can simulate multi-line intensity maps relatively quickly, which is useful for interpreting the signals once the observation of a particular line is made. This package also comes with several analysis techniques for calculating the three-dimensional isotropic power spectrum ($P_{3D} \left(k \right)$), the anisotropic power spectrum ($P_{3\mathrm{D}}\left(k_{\parallel}, k_\perp \right)$) in 3D, and the angular power spectrum in 2D ($C_\ell$), so that line intensity maps can be analyzed in different ways to extract the maximum encoded information. Simulations of multi-line intensity maps across a broad redshift range are helpful in performing cross-correlations between two line intensity maps at the same redshift, as they probe the same sources and underlying dark matter density fluctuations. Furthermore, scanning the Universe at different redshifts with the MLIM technique is not only a promising probe but also carries an opportunity to perform cross-correlations with galaxy surveys and CMB secondary anisotropies, e.g., CMB weak lensing, thermal and kinetic Sunyaev--Zel’dovich (tSZ and kSZ, respectively) effects \citep{Sato-Polito:2020cil, Schaan:2021gzb, Chung:xc}.

This paper is structured as follows: in Section \ref{sec:theory}, we provide an overview of the theoretical framework for line intensities, discussing their connection to the SFR of galaxies. In Section \ref{Sec:halo-model}, we introduce the halo-model formalism used to calculate the power spectrum of line intensities at specific redshifts and present the results obtained through this approach. In Section \ref{sec:sims}, we showcase the simulation results by describing the steps taken to generate intensity maps, and we present the detectability of the CII\,158 signal in Section \ref{sec:detectability}. Finally, in Section \ref{sec:conclusion}, we summarize our findings and conclusions. By exploring the theoretical and simulation-based aspects of line intensity mapping, we provide insights into astrophysical modelling uncertainties and the potential for future observational efforts in this area.

Throughout this study, we assume a flat $\Lambda$CDM universe with cosmological parameters as defined by the Planck TT, TE, EE+lowE+lensing results \citep{P18:main}. In the rest of this paper, we denote atomic line emission by writing together the line name and its wavelength in micrometre, e.g. CII\,158. For molecular line emission from CO, we denote the lines with the upper rotational transition level to the lower level, e.g., CO\,(1-0). We followed the same naming convention in $\limpy$, and line names can be passed to calculate the necessary quantities.

\section {Theory of line intensity mapping} 
\label{sec:theory}
The rest-frame frequency of a particular line emission, $\nu_{\rm rest}$, at redshift $z_{\rm em}$ will be observed at present by an instrument with an observational frequency $\nu_{\rm obs}$, such that $\nu_{\rm obs}=\nu_{\rm rest}/(1 + z_{\rm em})$. An instrument can be designed to probe a bright line from a broad redshift range to understand the different physical processes at that time by selecting several frequency channels. For instance, the Epoch of Reionization Spectrometer (EoR-Spec) on FYST with the observational frequency from 220-410 GHz is set up to detect the CII\,158 line emission across the broad redshift range of 7.6 to 3.6 \citep{CCAT-prime2021}. 
\begin{figure}
\centering
\includegraphics[width=0.5\textwidth]{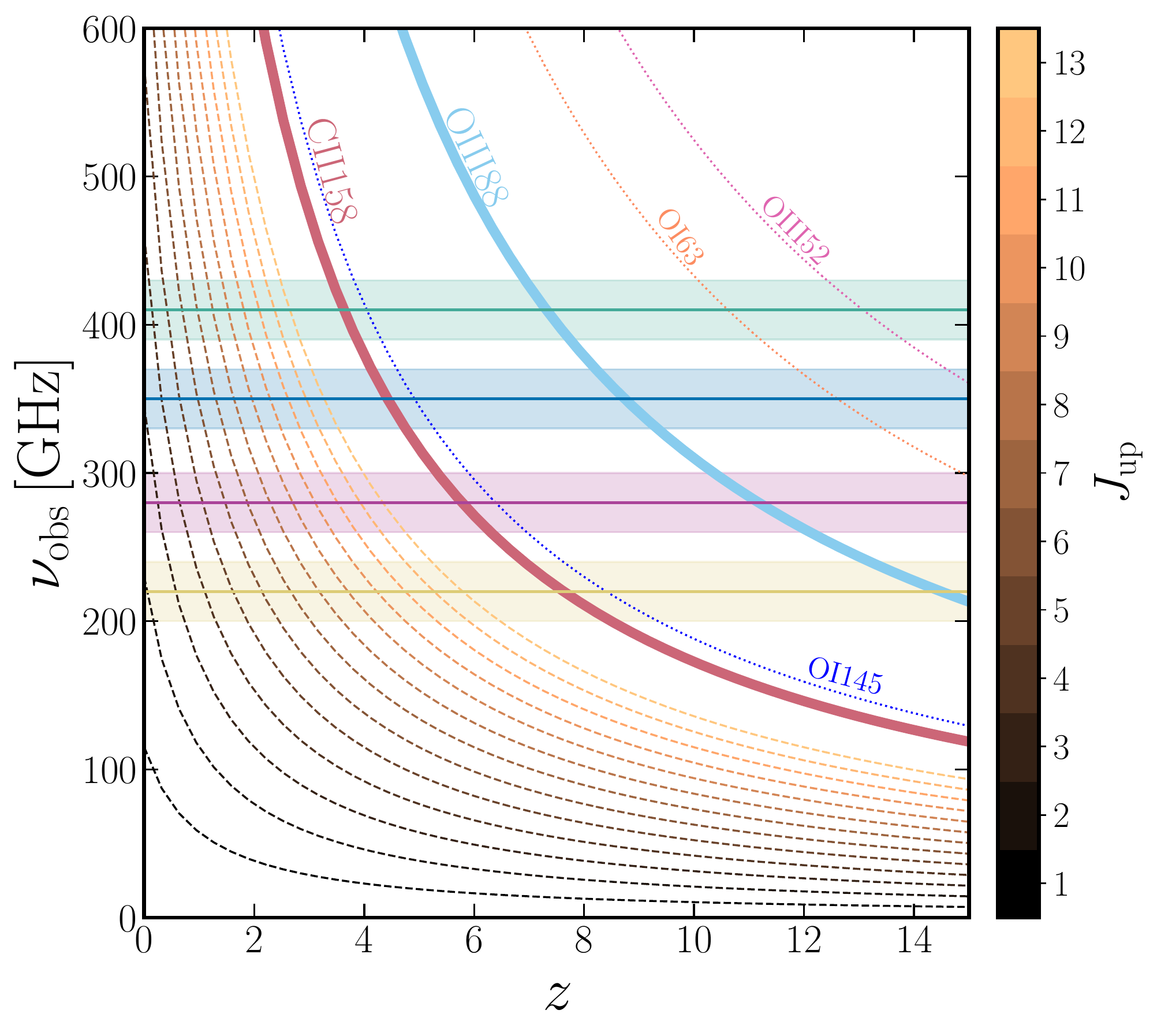} 
\caption{Redshift evolution of different atomic and molecular lines of interest in the redshift-frequency space. The four horizontal lines show the central frequencies of EoR-Spec on FYST, $\nu_{\rm cen}$, and the corresponding shaded regions represent the frequency bandwidths, $\Delta \nu$. Dashed lines show the rotational transitions of CO from the $J_{\mathrm{up}}$ to the $(J_\mathrm{up}-1)$ level.}
\label{fig:redshifted_lines}
\end{figure}

In Figure \ref{fig:redshifted_lines}, we show the redshift evolution of a few bright-line emissions that fall mainly in the FYST's EoR-Spec frequency coverage, 220 - 480\, GHz. All the lines that intersect the horizontal line representing a frequency channel will carry the information from the redshifts corresponding to the intersection. If EoR-Spec on FYST aims to detect the CII\,158 lines from $z\sim 7.3$ using $\nu_{\rm obs}=220$GHz, all other lines that cross the 220\, GHz frequency line, such as all CO transitions from CO\,(2-1) to CO\,(13-12), OIII\,88, and OI\,145, etc., will also come from different redshifts into the same frequency channel. In this case, one has to clean the signals of other lines to detect the desired line emission.
\begin{table}
\begin{center}
\begin{tabular}{cccc} 
\hline
\hline
$\nu_{\rm obs}$ & $z_{\mathrm{CII158}}$ & $z_{\mathrm{OIII88}}$ & $z_{\mathrm{CO76}}$ \\ \hline
220 & 7.6 & 14.5 &  2.66  \\
280 & 5.8 & 11.2 & 1.87 \\ 
350 & 4.4 & 8.7 & 1.30 \\
410 & 3.6 & 7.3 & 0.96  \\ 
\hline
\end{tabular}
\caption{Redshift of observation $z$ of CII\,158, OIII\,88, and CO\,(7-6) emission lines at different observational frequency bands $\nu_{\mathrm{obs}}$ of EoR-Spec on FYST \cite{CCAT-prime2021}.}
\end{center}
\end{table}\label{tab:fyst_freq}

The detection of CII\,158 and OIII\,88 line emissions at $z\gtrsim 6$  will play a crucial role in understanding the epoch of reionization. In contrast, detecting higher J-ladder transition will provide us with information about structure formation and galaxy evolution during the post-reionization era. To quantify their relative contributions to the total observed signal, we calculate the power spectra of the signals using both the halo model approach and N-body simulations. We show a few faint lines, such as OI\,145, OI\,63, OIII\,52, which could act as interlopers because of their redshift overlaps with FYST's EoR-Spec frequencies. Simple modelling of these lines is also important to understand the foreground contamination of CII\,158 and OIII\,88. The FYST's EoR-Spec survey will scan the sky with 
a frequency range of 220-410 GHz at a spectral resolution
of $R \sim 100$. This corresponds to the redshift coverage of $z_{\rm CII158} \sim 3.6 - 7.6$ for CII\,158 line and $z_{\rm OIII88}\sim 7.3 -14.5$. We show the redshifts for the line emission corresponding to a few central frequencies of FYST's EoR-Spec, such as 220\, GHz, 280 GHz, 350 GHz, and 410 GHz. With the $\limpy$ code, the intensities and power spectra of any selected lines can be generated at any redshift between $z\sim 0$-10. However, in this paper, we show the results only for the redshifts mentioned in Table\,\ref{tab:fyst_freq}.
\begin{figure*}
\includegraphics[width=\textwidth]{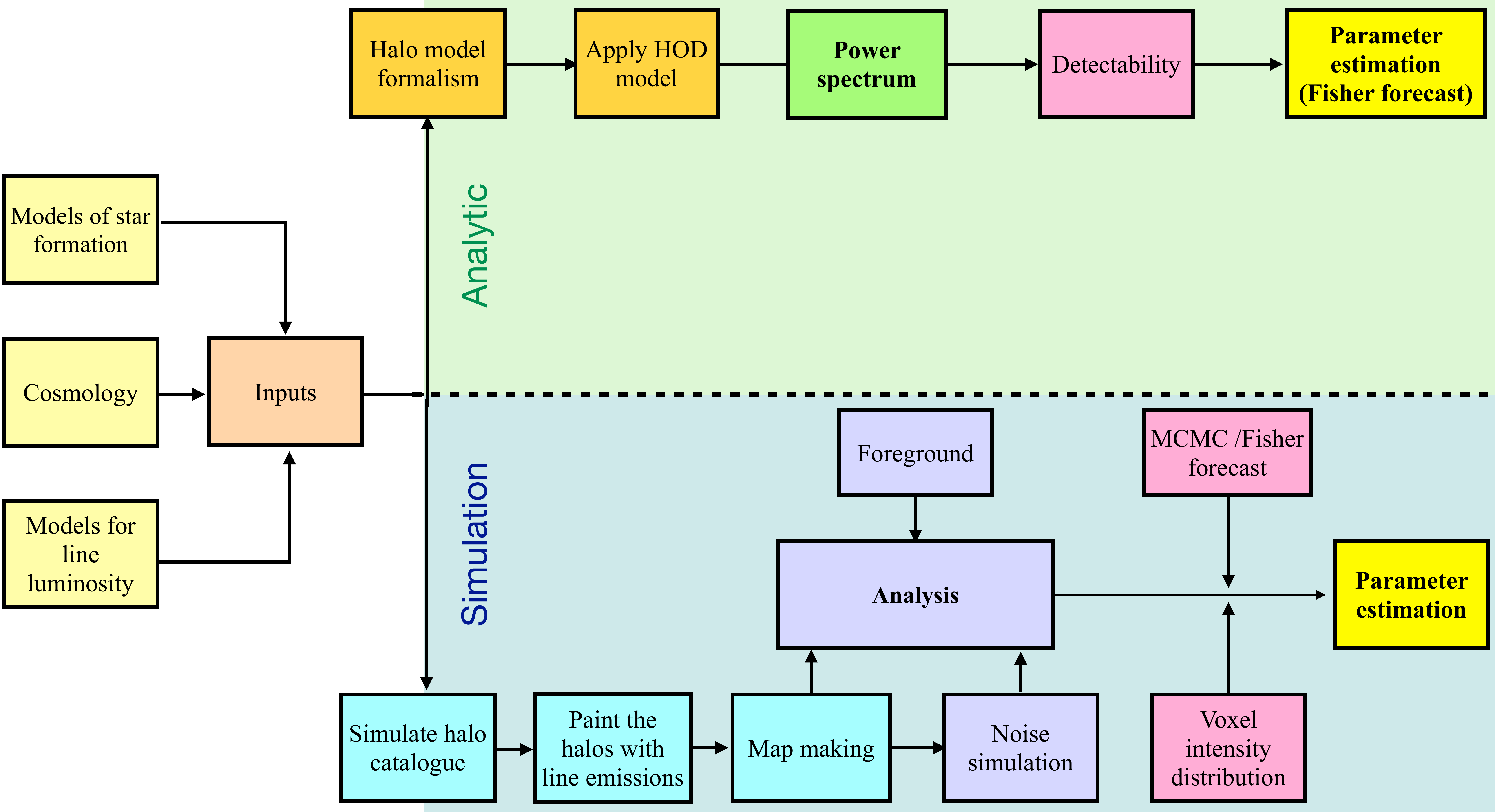} 
\caption{The schematic flowchart of the $\limpy$ package. Several built-in star formation models and multi-line luminosity models are implemented as inputs to the code. Based on these input choices, the package will calculate the power spectrum relying on either the halo model approach or painting the line luminosities on an externally provided halo catalogue. The package can make line intensity maps, and if the specification of an experiment is provided, it can calculate the signal-to-noise ratio. Furthermore, $\limpy$ can be used for parameter estimation based on Markov Chain Monte Carlo (MCMC) or Fisher matrix methods. These methods incorporate observational data to infer the parameters that best describe the underlying astrophysical processes. }
\label{fig:LIM_flowchart}
\end{figure*}

\begin{figure*}
\includegraphics[width=\textwidth]{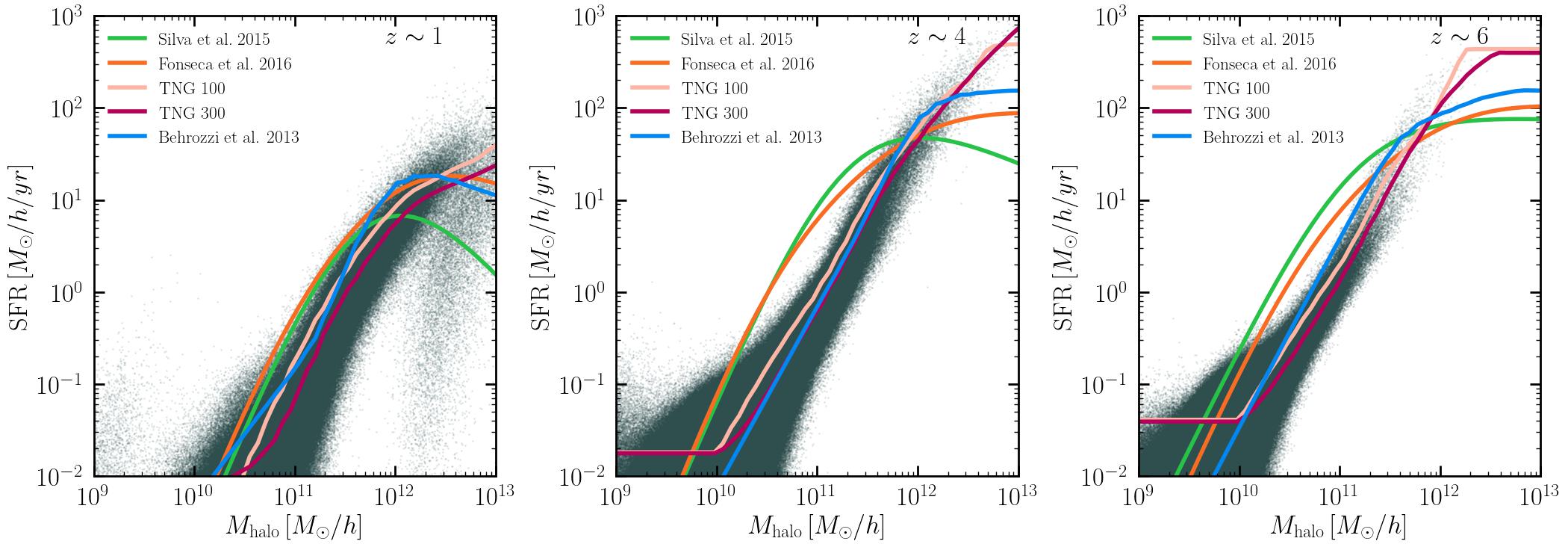} 
\caption{We illustrate the assumed SFR models as a function of halo mass, three different redshifts: $z\sim1$ (left), $z\sim4$ (middle), and $z\sim6$ (right). The scatter points represent the star formation histories of individual halos in the TNG300 simulations \citep{TNG-gen}, while the pink and purple solid lines depict the best-fit curves based on the TNG100 and TNG300 simulations. For comparison, we show the interpolated SFR from \citet{Behroozi2019}, and analytic models of SFR taken from \citet{Silva:2015} and \citet{Fonseca:2016}. This plot captures the complex picture of star formation history in halos, as all dark matter halos with the same mass do not form the same amount of stars. The uncertainty in the SFR can propagate to the luminosity of the various emission lines. Careful consideration of the uncertainties in the SFR is necessary when interpreting intensity mapping observations and making predictions about the underlying astrophysical processes that drive the formation and evolution of galaxies over cosmic time. }
\label{fig:SFR_mhalo}
\end{figure*}

We describe the workflow of the $\limpy$ package in Figure \ref{fig:LIM_flowchart}. The main ingredients are fed to the code as input to initialize it. In Section \ref{sec:sfr} we describe the in-build models of star formation histories of galaxies that can be passed to the code by mentioning the model name according to the documentation. The default cosmological parameters are based on the Planck 2018 paper. Users can modify these parameters by modifying the input file. There are several models that convert the SFR to the different line luminosities, and these models can be passed by changing the inputs. Once the basic cosmological and astrophysical parameters are initialized, then the code can calculate the power spectrum either based on the halo-model approach or simulations. Next, we calculate the power spectrum and forecast the signal-to-noise ratio for a particular experiment, providing the configurations of a telescope for the white noise calculation. In principle, different noise sources, such as atmospheric, foreground contamination, instrumental white noise, etc., can be passed in the code altogether to calculate the signal-to-noise ratio. The final goal is to make forecasts for parameters based on particular observations or mock data. Either of Fisher forecasts or MCMC algorithms can be applied for parameter estimation using the \textsf{$\limpy$} modules. 

In the following subsections, we review and summarize the basic properties of star formation histories and their relationship with atomic and molecular line luminosities. These models are based on several assumptions, and changing those assumptions will typically lead to a change in results. The detailed analysis and interpretation of SFR based on galaxy formation models are out of the scope of this paper. Our main goal is to quantify the SFR as a function of halo mass $M_{\rm halo}$ and $z$ so that we can calculate the necessary quantities to estimate the power spectra of line emissions over a broad redshift range.

\subsection{Empirical models of SFR}\label{sec:sfr}
One of the most complex problems in the field of modern astrophysics is how stars form in galaxies and their role in the process of galaxy evolution. The SFR across cosmic time and its relation with halo mass are key to understanding galaxies' morphology,  chemical and physical properties. Several simulation suites of galaxy formation incorporate complicated astrophysical processes in galaxies and are capable of shedding light on the SFR$- M_{\rm halo}$ relation across cosmic time \cite[e.g.,][]{Eagle-sim-2015, TNG-gen, Fable-sim-2018,  Behroozi2019}. Multi-wavelength observations of galaxies in the UV by \textit{HST} and Far-IR observations by the \textit{Herschel} telescope reconstructed the cosmic star formation density out to redshift $z\lesssim 10$ \citep{Madau:2014bja}. 
Due to the lack of observational data at high redshift ($z \gtrsim 4$), statistical errors on the SFRD increase significantly. We aim to reconstruct the SFR empirically from several models and simulations that vary the mass of host halos across a wide range of redshifts. We incorporate five SFR models that can be used to produce line intensity maps, namely \texttt{Behroozi19} from $\universemachine $\citep{Behroozi2019}, \texttt{Tng300} and \texttt{Tng100} \citep{TNG-gen, TNG-gal} from $\tng$, as well as fitting functions such as \texttt{Silva15} \citep{Silva:2015} and \texttt{Fonseca16} \citep{Fonseca:2016}. 

In \texttt{Silva15} \citep{Silva:2015}, the average SFR is extracted from the post-processed simulated galaxy catalogue \citep{DeLucia2006, Guo2011}, in which the minimum halo mass is set to $10^8\,M_\odot$/$h$. The $\mathrm{SFR}-M_{\rm halo}$ scaling relation is applicable over a broad redshift range, from $z=0$ to 20. This empirical relation can be approximated to $z \lesssim 20$ for studying the high redshift line intensities, particularly for OIII\,88 and OIII\,52. In the \texttt{Silva15} SFR model, SFR is parameterized as a function of two power-law terms of halo mass, whereas the \texttt{Fonseca16} model uses the SFR based on the same simulated catalogue as the three power law exponent of halo masses. The parameters for the SFR function according to the \texttt{Fonseca16} model are given for the redshift range 0-10, and we keep the SFR fixed for the redshift range $10-20$ as the same as for the SFR at redshift $z=10$. Various physical processes, such as galaxy mergers, the effect of the environment, feedback, etc., are involved in the SFR across a broad range of redshifts, which is very complex to model. Hence, we adopt fitting SFR functions such as \citet{Fonseca:2016} and \citet{Silva:2015}. For comparison, we use the output of SFR for each halo from IllustrisTNG simulations done in box size for $L=100\,$ cMpc and $300\,$cMpc (hereafter TNG100 and TNG300, respectively) \citep{TNG-gen, TNG-gal}. 
%%%%%%%%%%%%%%%%%  

We adopt the output of the $\universemachine$ simulations\footnote{\href{https://www.peterbehroozi.com/data.html}{https://www.peterbehroozi.com/data.html}} to infer the SFR across the redshift $z=0-10$ \citep{Behroozi2019}. The empirical methods for tracking down the SFR of each halo across the redshift range are constrained by the observations such as galaxy UV luminosity functions, observed stellar mass functions, quenched fractions, etc. Furthermore, to evaluate the best-fit function from the scattered SFRs from the \texttt{TNG100} and \texttt{TNG300} simulations, we take 100 bins in halo mass from $10^{10}\,M_\odot/h$ to $10^{15}\, M_\odot/h$ and take the median value of all the SFRs that fall into each mass bin. We fix the minimum mass of the line emission sources to $M_{\rm min}= 10^{10}\, M_\odot/h$ throughout this paper. We use these star formation histories to understand the multi-line luminosities and their astrophysical uncertainties due to the scatter of SFR for a fixed halo mass. 

In Figure \ref{fig:SFR_mhalo}, we show how the SFR varies with the masses of host dark matter halos as predicted by different empirical models. The ratio between the maximum and minimum SFR for a fixed halo mass varies at different redshifts due to the complex physical process of star formation. For $M_{\rm halo}=10^{11}\, M_\odot/h$, this ratio becomes 250, 96, and 11 at redshifts $z\approx 1$, 4, and 6, respectively. This figure provides insights into the evolution of star formation in halos of varying masses and at different cosmic epochs, shedding light on the complex interplay between dark matter, gas, and other astrophysical processes that shape the growth and evolution of galaxies over time. The exact nature of the $\mathrm{SFR}-M_{\rm halo}$ relationship is complex and depends on several factors, including the efficiency of gas cooling, the ability of gas to collapse into small-scale structures, and feedback processes such as supernova explosions that can regulate star formation \citep{Conroy2009}. Understanding the underlying physical mechanisms that govern the star formation process in galaxies and their dependence on halo mass and redshift is crucial for developing a comprehensive picture of galaxy formation and evolution \citep{Limfast2}. For simplicity and optimization purposes, we use the average $\mathrm{SFR}- M_{\rm halo}$ relation to estimate the multi-line luminosities, ignoring the dependencies of other astrophysical parameters related to the complex star formation history in haloes. 

\begin{figure*}
\includegraphics[width=\textwidth]{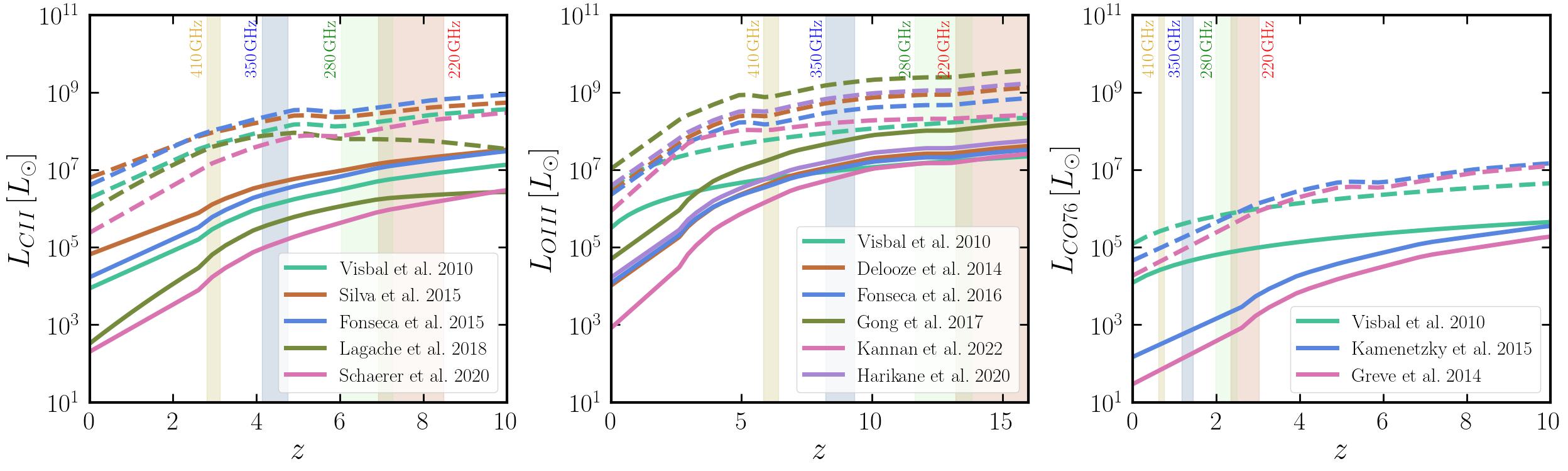} 
\caption{Redshift evolution of the CII\,158, CO76, and OIII\,88 luminosities based on the models mentioned in the legends. Solid and dashed lines are the line luminosities for the halo mass $10^{10}\,M_\odot/h$ and $10^{11}\,M_\odot/h$, respectively. The vertical-shaded regions are the redshift coverage of these lines corresponding to the frequency bandwidths of the EoR-Spec on FYST.}
\label{fig:line_luminosities}
\end{figure*}

\subsection{SFR - $L_{\rm line}$ relation}
A crucial question that arises in the development of Line Intensity Mapping (LIM) models is the identification of the key factors that trace the observed multi-line luminosities in galaxies. It is assumed that multi-line luminosities trace the star formation histories, and SFR can be converted to line luminosities using a power-law relation. In the previous subsection, we modelled how the SFR depends on the mass of halos, and we discussed how the multi-line luminosities are related to the SFR so that for a given halo mass, we can estimate the multi-line luminosities. We incorporated several $L_{\rm line}-M_{\rm halo}$ relations in $\limpy$ to study the modelling uncertainties in the intensity maps. 

The scaling relations for $L_{\rm CII158}-SFR$ can be called by name as \texttt{Visbal10} \citep{Visbal2010}, \texttt{Silva15-m1}, \texttt{Silva15-m2}, \texttt{Silva15-m3}, \texttt{Silva15-m4} \citep{Silva:2015}, \texttt{Fonseca16} \citep{Fonseca:2016}, \texttt{Lagache18} \citep{Lagache2018}, \texttt{Schaerer20} \citep{Schaerer_2020}. The luminosity of these lines scales with the SFR as $L_{\rm line} = R_{\rm line} \times\,\mathrm{SFR}$ for \texttt{Visbal10} model, where $R_{\rm line}$ is the conversion factor \citep{Visbal2010}, which does not evolve with the redshifts. Assuming all galaxies have the same $R_{\rm line}$, their values are given by $6\times 10^6$ and $2.3\times 10^6$ in $L_\odot/(M_\odot/yr)$ unit for CII\,158 and OIII\,88 lines, respectively. The values of $R_{\rm line}$ for J-ladder transitions of CO molecules are given in the \citealt[see Table 1]{Visbal2010}. In the
\texttt{Silva15-m1} (and m2, m3, and m4) model, the luminosity of CII\,158 is modelled as a power law relation; $\log L_{CII\,158}= \alpha + \beta\log(SFR)$. The four sets of models are given by the different values of $\alpha$ and $\beta$ that we specify in $\limpy$ by the name 
\texttt{Silva15-m1},\texttt{ Silva15-m2}, \texttt{Silva15-m3}, and \texttt{Silva15-m4} \citep{Silva:2015}. For the \texttt{Fonseca16} model, the luminosity of CII\,158, OIII\,88, OI\,145, OI\,63, and OIII\,52 lines can be expressed as the same power law relation with SFR, but the coefficients are different than \texttt{Silva15} model as mentioned in \citet{Fonseca:2016}. For the \texttt{Silva15} and \texttt{Fonseca16} models, the coefficients $\alpha$ and $\beta$ do not change with the redshift, but the multi-line luminosity of lines varies with the halo mass only because of the evolution of SFR with redshift. The redshift evolution of the coefficients is captured with a modified version of the power in the \texttt{Lagache18} model \citep{Lagache2018}. In addition, we also model the scaling relation from low redshift observations of the molecular line from the ALMA-Alpine experiment that can be written in a similar form. The mean values of $\alpha$ and $\beta$ are given in Table 2 of \citet{Schaerer_2020}. 

For modelling OIII\,88 lines, we include the $L_{\rm OIII88}-SFR$ scaling relation defined in the code by \texttt{Visbal10} \citep{Visbal2010}, \texttt{Delooze14} \citep{DeLooze:2014}, \texttt{Fonseca16} \citep{Fonseca:2016}, \texttt{Gong17} \citep{Gong2017}, \texttt{Harikane20} \citep{Harikane2020}, and \texttt{Kannan21} \citep{Kannan:2021ucy}. The SFR in the far-infrared is modelled in terms of the CII\,158, OI\,63, and OIII\,88 line emissions from the Herschel Dwarf Galaxy Survey, and the scaling relations are obtained from \citealt[see Table 2]{DeLooze:2014}. They find that OI\,63 and OIII\,88 trace the SFR better than the CII\,158 lines and the disperse in the relation between SFR and $L_{\rm line}$ for OIII\,88 and OI\,63 is improved by a factor of $\sim 2$ compared with the CII\,158 lines. In addition to that, we adopt the scaling relations between SFR and luminosity of OIII\,88 lines based on the observations by ALMA at $z\sim 6-9$ \citep{Harikane2020}. They find the ratio $L_{\rm OIII\,88}/L_{\rm CII\,158}$ can be 10 times higher than this ratio at $z\sim 0$, suggesting a strong redshift evolution of line luminosities. While we implement this scaling relation in $\limpy$, the OIII\,88 line luminosities can change across redshift due to the change of SFR. Furthermore, we use the scaling relation of OIII\,88 that is derived from the observed luminosity function and SFRD at $z \lesssim 5$ \citealt[see Section 2]{Gong2017}. 

We use several scaling relations to model CO line emissions in our study. One such relation is based on the spectra obtained from the \textit{Herschel} SPIRE Fourier Transform Spectrometer \citep{Kamenetzky2016}. The luminosity of CO molecular emission, $L_{\rm CO}$, is found to depend on the FIR luminosity of galaxy samples $L_{\rm FIR}$. In order to calculate the luminosity of all lines for a given SFR, we convert it into $L_{\rm FIR}$, where $L_{\rm FIR} = 1.1 \times 10^{10} \times SFR$ \citep{Carilli2011}. Additionally, we incorporate another model for CO molecular transitions based on ALMA observations \citep{Greve2014}. This model allows us to estimate the luminosity of full rotational transitions of CO molecules using data from \textit{Herschel} SPIRE-FTS and ground-based telescopes \citealt[see Table 3]{Greve2014}. The full J-ladder rotational transitions of CO are essential for understanding the relative contributions in the interlopers which contaminate the desired signal that we aim to detect. Using multiple models allows us to estimate the CO line emissions, which helps us to account for the uncertainties associated with these relations.

\begin{figure*}
\includegraphics[width=1\textwidth]{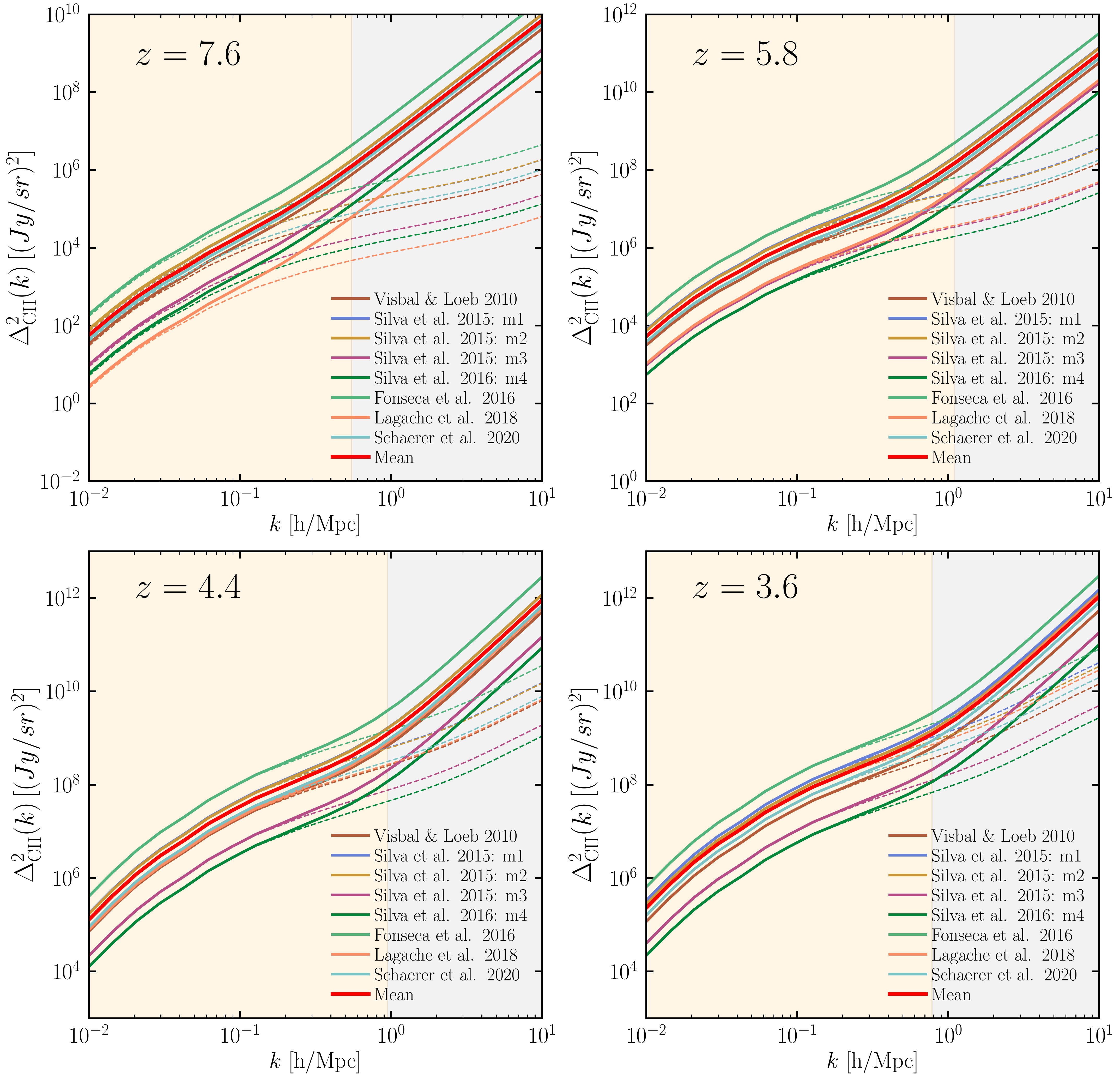} 
\caption{The power spectra of the CII\,158 line at redshifts 7.6, 5.8, 4.4, and 3.6. The dashed lines in each panel show the contribution to the signal from the clustering term alone, while the solid lines represent the total signal, including both the clustering and shot noise terms. The solid red lines in each panel are the mean line of all the models, representing the average signal predicted by the different theoretical models that we consider here. The shaded grey region represents the scales where the dominant term is shot noise, while the yellow-shaded area roughly corresponds to the scales where the clustering term is larger than the shot noise term. The shape and amplitude of the CII\,158 have important implications for studying ISM physics and structure formation at high redshift, as they provide insights into the large-scale structure of matter in the Universe and the properties of the sources that generate the CII\,158 signal.}
\label{fig:CII_powerspectrum}
\end{figure*}

Figure \ref{fig:line_luminosities} shows the evolution of different line luminosities for the \texttt{Silva15} star formation model. The plot shows the ratio between the maximum and minimum luminosities of the CII\,158 lines, the CO\,(7-6) line, and the OIII\,88 line as a function of redshift for a fixed minimum halo mass of $M_{\rm min}=10^{10}\, M_\odot/h$. At a redshift of $z\sim 3.8$, the ratio between the maximum and minimum CII\,158 line luminosities is approximately 45. However, for the same minimum halo mass, the ratio for CO\,(7-6) luminosity becomes 110 at $z\sim 2.66$ and 310 at $z\sim 0.96$, highlighting the increasing spread in luminosity ratios as the redshift decreases. The ratio for the CO\,(7-6) line decreases to 1.5 and 6, respectively, if the minimum halo mass is set to $10^{11}\, M_\odot/h$, suggesting that the choice of minimum halo mass can significantly impact the luminosity ratios. Finally, for the OIII\,88 line luminosity with $M_{\rm min}=10^{10}\, M_\odot/h$, the ratio between the maximum and minimum luminosity is 6 and 8 at redshifts z= 14.5 and 7.3, respectively, indicating that this line is less sensitive to changes in redshift than the other lines considered in the plot.

\section{Analytic model}\label{Sec:halo-model}
In this section, we employ a halo model formalism to compute the power spectrum of multi-line intensities. The intensity of lines emitted at $z_{\rm em}$ can be expressed as
\begin{equation}
I_{\rm line}(z)= \frac{c}{4\pi}\frac{1}{{\nu_{\rm rest}}H(z_{\rm em})} \int_{M_{\rm min}}^{M_{\rm max}} L_{\rm line}(M,z)\frac{dn}{dM}dM\,.
\end{equation}\label{eq:Iline}
In this equation, $c$ represents the speed of light in a vacuum, and $H(z_{\rm em})$ denotes the Hubble parameter at the redshift of line emission. The halo mass function is represented by $dn/dM$. Throughout our study, we utilize the Tinker halo mass function for our calculations \citep{Tinker:2008ff}. Here, $M_{\rm min}$ refers to the minimum mass of the halos contributing to the intensity maps, while $M_{\rm max}$ signifies the upper mass threshold of the sources.

The power spectrum of fluctuations of the line intensity is the summation of 1-halo and 2-halo terms that can be written as 
\begin{equation}
P_{\rm line}(k,z) = I_{\rm line}(z)^2 \left[b_{\rm line} (z) P_m(k, z) + P^{\rm shot}_{\rm line} (z) \right]. 
\label{1q:pline_halo}
\end{equation}
In the above equation, $P_m(k,z)$ is the matter power spectrum and $P_{\rm shot}$ is the shot noise term. We calculate the matter power spectrum using CAMB \citep{CAMB} under the linear approximation. The bias of the line emission, $b_{\rm line}$, is proportional to the bias of line emitting sources, which can be written as
\begin{equation}
b_{\rm line}(z) = \frac{\int_{M_{\rm min}}^{M_{\rm max}} dM (dn/dM) L_{\rm line}(z) b_h (M, z)}{\int_{M_{\rm min}}^{M_{\rm max}} dM (dn/dM) L_{\rm line}(z)}\,.
\label{eq:bline}
\end{equation} 
Here, $b_h$ is the bias of dark matter halos. We use Colossus\footnote{\href{https://bdiemer.bitbucket.io/colossus/}{https://bdiemer.bitbucket.io/colossus/}} package to calculate the bias of dark matter halos and halo mass function \citep{Colossus}. The bias of lines accounts for the clustering properties of the power spectrum, which has significant contributions at large scales. 

Finally, the shot noise term of the power spectrum is proportional to the line luminosity function, which is given by
\begin{equation}
P^{\rm shot}_{\rm line}(z) = \frac{\int_{M_{\rm min}}^{M_{\rm max}} dM (dn/dM) L_{\rm line}(z)^2}{\left[\int_{M_{\rm min}}^{M_{\rm max}} dM (dn/dM) L_{\rm line}(z)\right]^2}\,.
\label{eq:pshot}
\end{equation}
The shot noise term has the same contributions at all scales. 

Figure \ref{fig:CII_powerspectrum} presents the power spectrum of CII\,158 for various models, highlighting the clustering and total signal comprising both the clustering and shot noise terms. Despite using the same SFR as input, the $L_{\rm line} - SFR$ relation, the dispersion on the amplitude of power spectra at these redshifts due to the modelling differences exceed one order of magnitude at these redshifts. Some models are based on the $L_{\rm line} - M_{\rm halo}$ relation, while others convert SFR to CII\,158 line luminosities. For the latter case, we first calculate the SFR for different halo masses and then determine the line luminosities based on the $L_{\rm line} - SFR$ relation. Similarly, we can generate the power spectra of OIII\,88 and molecular lines from CO\,(1-0) to CO\,(13-12) for available models. For an example case, we show the power spectra of OIII\,88 and CO\,(7-6) at redshifts corresponding to FYST's EoR-Spec in Appendix \ref{sec:app_power}. In this way, we can assess the contribution of these lines to the total signal at a particular frequency channel and determine their detectability in the presence of interlopers.

In Equations (\ref{eq:bline}) and (\ref{eq:pshot}), we made the assumption that there is only one star-forming source in each halo. However, this is not the case for high-mass halos. To account for multiple star-forming sources in halos, we employ the halo occupation distribution (HOD) model. The mean occupation functions for central and satellite galaxies in a halo of mass $M_h$ are given by \citep{Zheng2004}:
\begin{equation}
\langle N_{\rm cen}(M_h) \rangle=\frac{1}{2}\left[ 1+ \mathrm{erf}\left(\frac{\log M_{\rm h}-\log M_{\rm th}}{\sigma_{\rm logM}}\right)\right]\,,
\label{eq:N_cen}
\end{equation}
\begin{equation}
\langle N_{\rm sat}(M_h) \rangle=\left( \frac{M_{\rm h}-M_{\rm cut}}{M_1}\right)^{\alpha_{\rm g}} \,.
\label{eq:N_sat}
\end{equation}
In the above equations, $\langle N_{\rm cen}(M_h) \rangle$ and $\langle N_{\rm sat}(M_h) \rangle$ represent the average number density of central and satellite galaxies, respectively. $M_{\rm th}$ denotes the threshold halo mass required to host a central galaxy, while $M_{\rm cut}$ represents the minimum mass necessary for hosting satellite galaxies. $\sigma_{\rm logM}$ is the width of the transition in the step-like error function, $\alpha_{\rm g}$ refers to the power law exponent, and $M_1$ is the mass normalization factor. The HOD-model parameters are given as $\log M_{\rm th}=10^8$, $\sigma_{\rm logM}=0.287$, $\log M_{\rm cut}=12.95$, $\log M_{1}=13.62$, and $\alpha=0.98$ \citep{Zheng2004}. By incorporating the HOD model, we can more accurately account for the distribution of star-forming sources in halos, particularly in high mass halos. The inclusion of the HOD model provides a more comprehensive picture of the power spectrum of multi-line intensities.

\begin{figure}
\includegraphics[width=0.5\textwidth]{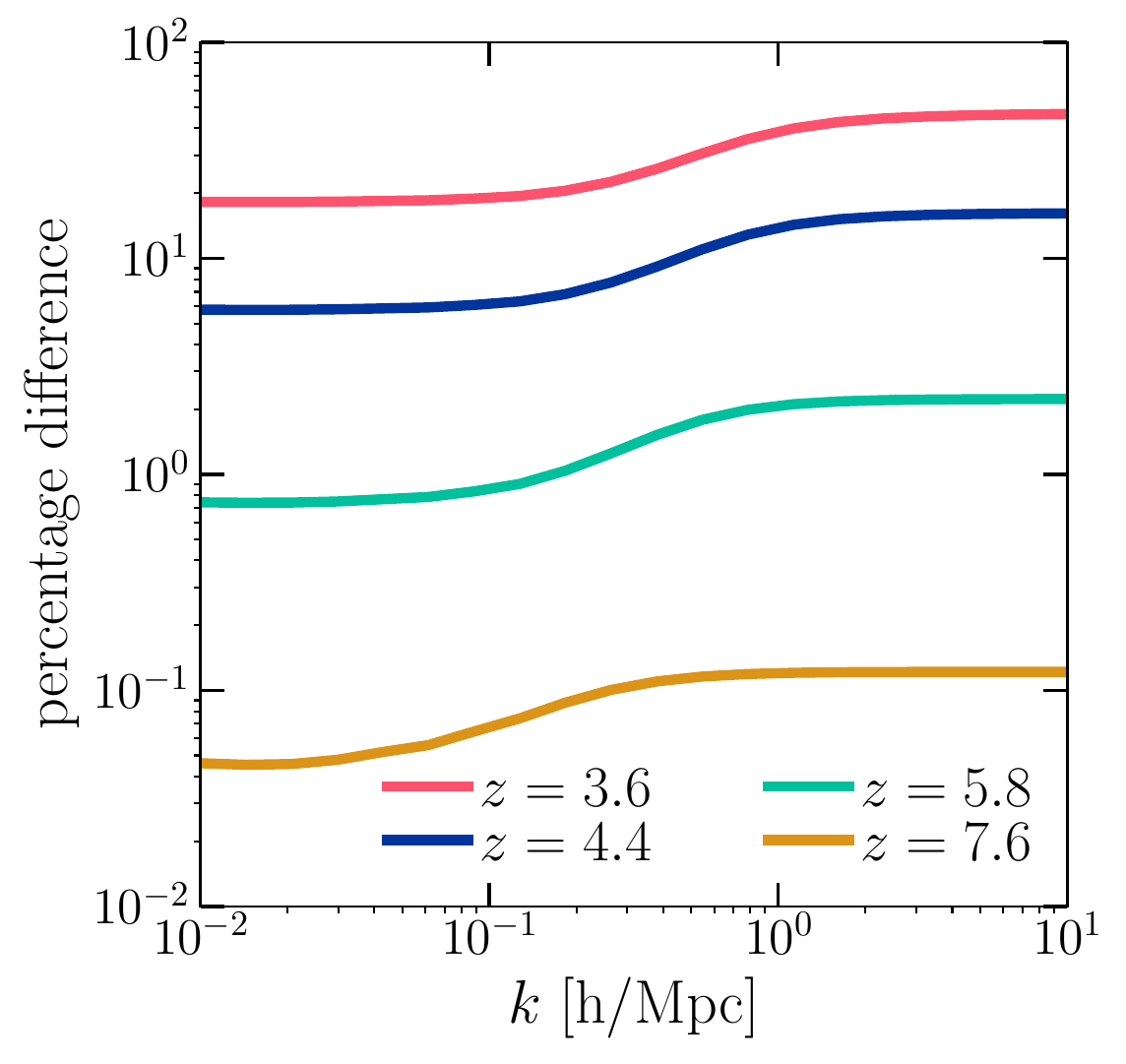} 
\caption{Percentage difference in the CII\,158 power spectrum due to the use of a HOD model. For this example, we calculate the power spectrum based on the \texttt{Silva15} star formation model and \texttt{Fonseca16} line luminosity model of CII\,158. It shows that the inclusion of the HOD model is more important for the power spectrum of CII\,158 at low redshift.}
\label{fig:HOD_diff}
\end{figure}

Figure \ref{fig:HOD_diff} presents the percentage difference in the power spectra of CII\,158 lines resulting from the inclusion of the HOD model in our calculations. The HOD model accounts for the line emissions both from central and satellite galaxies, and its inclusion can significantly affect the 1-halo term of power spectra. The plot shows the increase in the power spectra due to the HOD model for different redshifts and scales, represented by the percentage difference compared to the power spectra without HOD. At a scale of $k\sim 5$\, $h\,\mathrm{Mpc}^{-1}$, the power spectra of CII\,158 lines increase by 73\%, 25\%, 2\%, and 0.1\% at redshifts 3.6, 4.4, 5.8, and 7.6, respectively, highlighting the significant impact of the HOD model on the power spectra of CII\,158  lines.

\begin{figure*}
\includegraphics[width=\textwidth]{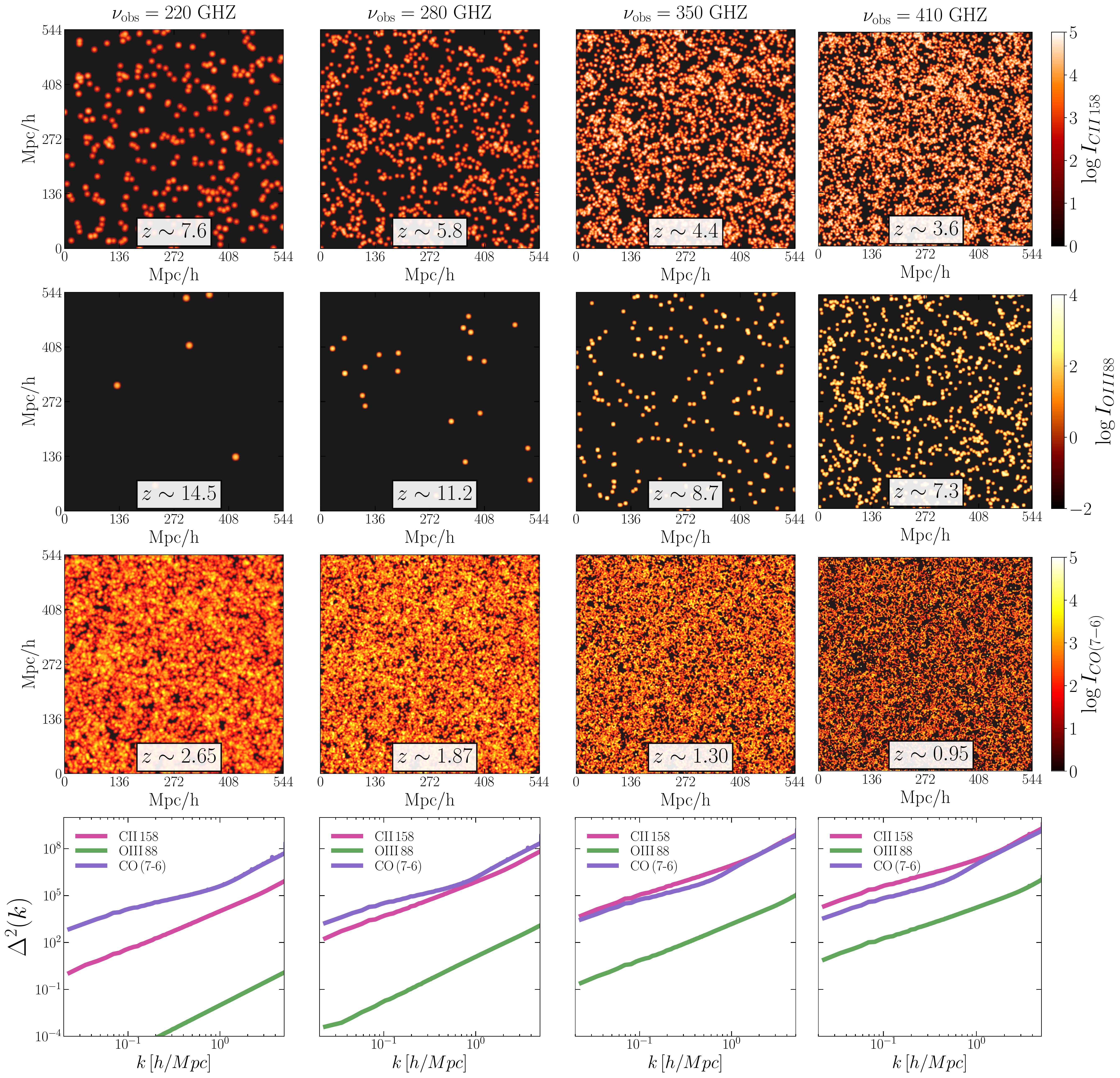}
\caption{We display simulated maps of CII\,158 (first row), OIII\,88 (second row), CO (7-6) (third row) line intensities at redshifts corresponding to the central frequencies of the FYST's EoR-Spec experiment. The simulation boxes are generated from \texttt{21cmFAST} package for a 544\,$c$Mpc/$h$ box, and we keep fixed the same initial condition for all the simulations at different redshifts. The columns correspond to the specific observational frequency, as indicated by the column titles. The fourth row of the plot compares the dimensionless power spectra of these lines based on the \texttt{Tng300} star formation model and the \texttt{Visbal10} line luminosity model \citep{Visbal2010}. For visualization purposes, Gaussian beam convolutions were applied to the maps with full-width half-maximum (FWHM) beam sizes of 58, 45, 37, and 32 arc seconds from left to right.} 
\label{fig:simulated_maps}
\end{figure*} 

\section{Simulated maps of MLIM} \label{sec:sims}
In addition to theoretical modelling, $\limpy$ also performs the simulation of multi-line intensity maps at several redshifts. We utilize the semi-numerical cosmological simulation, 21cmFAST\footnote{\href{https://github.com/andreimesinger/21cmFAST}{https://github.com/andreimesinger/21cmFAST}}, to generate the dark matter halo catalogue. We execute the simulation on a box with a length, L= 800 $c$Mpc ($\approx$ 544 $c$Mpc/$h$) \citep{21cmfast}. The initial conditions for generating the perturbations in density are set at $z=300$. The density field evolves over cosmic time following linear perturbation theory. Next, we generate snapshots of several redshifts for different lines corresponding to the FYST's EoR-Spec frequency channels. We set the minimum mass of the halos to $M_{\rm min}=10^{10}\,M_\odot/h$. The simulation setup uses a number of grids along the box length, $N_{\rm grid}=1024$, meaning that the total number of dark matter particles is $N_{\rm grid}^3$. This enables us to step down to a length of $0.53$ $c$Mpc/$h$. After obtaining the halo field or resolved catalogues from simulations, we assign a specific line luminosity to those halos based on an SFR model and line luminosity model.

The primary advantage of line intensity mapping simulations is that we do not need to resolve individual sources. Consequently, we use a low-resolution simulation with a smaller $N_{\rm grid}$. This approach reduces the simulation time. We save all the intensity grids at different redshifts to calculate the power spectrum.

The 3D line intensity power spectrum in a simulation box is expressed by:
\begin{equation}
\Delta^2_{\rm line}(k) = \frac{1}{V_{\rm box}} \frac{k^3}{2\pi^2} \langle \tilde{I}^2(k)\rangle
\label{eq:pk_sims}
\end{equation}

Here, $V_{\rm box}$ represents the total volume of the simulation box, and $\tilde{I}$ is the Fourier transform of the intensity grid. We then perform the Fourier transform of the intensity grid using the NumPy FFT module. The intensity of each cell is calculated as \citep{Dumitru2018}:

\begin{equation}
I_{\rm cell} = \frac{c}{4\pi} \frac{1}{\nu_{\rm rest} H(z_{\rm em})} \frac{L_{\rm line, cell}}{V_{\rm cell}}\,.
\end{equation}

In principle, CO transitions with $J_{up}\geq 4$ at low redshifts will act as interlopers for all four of FYST's EoR-Spec frequency channels. However, we only show the CO\,(7-6) transition as an example case. We project the intensity grid of the length of 35 Mpc, 16 Mpc, 60 Mpc for CII\,158, CO\,(7-6), and OIII\,88 lines, which roughly correspond to the frequency resolution of the EoR-Spec on FYST at the central frequency, 280\, GHz. 

Once we generate the intensity grid with a reasonable value of $N_{\rm grid}$ to achieve the $M_{\rm min}$ for line emitting sources, we need to incorporate the effect of frequency resolution ($\delta \nu_{\rm obs}$) as an experiment cannot resolve the sources along the redshift axis for the smaller value than the $\delta \nu_{\rm obs}$. Therefore, the effective number of grids along the redshift axis will depend on the frequency resolution of an experiment. With a resolution of $N_{\rm grid}=1024$, we select halos above the mass $M_{\rm min} \gtrsim 10^{10}$ $M_\odot$/$h$. The 21cmFAST code does not explicitly resolve each halo in the simulation box, but it generates a halo field semi-numerically that can be accurately compared with the output from N-body simulations \citep{21cmfast, Limfast1}. This process saves memory usage and makes it run faster. If $\delta \nu_{\rm obs}$ is 2.8\,GHz around the central observational frequency, $\nu_{\rm obs}=220\,$GHz for an experiment to probe CII\,158 line emission, the corresponding redshift resolution is $\delta z \approx 0.07$. This $\delta z$ corresponds to the box length along the redshift axis (we define it to be along the $z$ axis of the cartesian coordinate system) $\delta L_z \approx 44\,c$Mpc/$h$ and $\delta N_{grid, z}\approx 90$. Therefore, the experiment with this configuration cannot resolve the sources along the redshift axis that fall between $z = 5.80$ and 5.87. In this case, the total number of cells of the same simulation box becomes $1024 \times 1024 \times 11$ as the grid points along the $z$-axis reduces to $N_{grid, z}=1024/\delta N_{grid, z} \approx 11$. We take the average intensity of all the intensity grids within this frequency resolution. We note if there is no mention of $\nu_{\rm obs}$ or the length corresponding to $\nu_{\rm obs}$ exceeds the length of the simulation box along the z-direction, the code does not apply the effect of the frequency resolution and calculates the power spectrum based on the grid points that were used to generate the halo catalogue. 

The code presented in this paper is versatile and can accommodate any type of halo catalogue that is provided as input. Our code requires the halo catalogue to contain two essential pieces of information: the halo mass in units of $M_\odot$/$h$, and the halo positions $(x, y, z)$ in Cartesian coordinates specified in units of Mpc/$h$. By accepting any halo catalogue as an input, the code offers the flexibility to utilize halo catalogues generated from full N-body simulations and perform detailed astrophysical analyses. This functionality is especially useful in studying the properties and evolution of halos in large-scale structure simulations, as well as in exploring the connection between halo properties and other astrophysical observables. 
%The ability to use any halo catalogue allows the $\limpy$ code to be easily integrated into existing simulation pipelines, providing a flexible tool to study intensity mapping in the context of complex galaxy formation physics.

Next, we apply beam convolution techniques to recreate the actual observation. Although the actual beam of an experiment can have a complex pattern, for simplicity we assume the beam can be approximated to be Gaussian. Then the beam pattern is characterized by $\theta_{\rm FWHM}$, the beam size in arcminute unit at the full width at half maximum (FWHM), and the standard deviation of the beam is $\sigma_{\rm beam} = \theta_{\rm FWHM}/ \sqrt{(8 \log 2)}$. We used the Astropy\footnote{\href{https://www.astropy.org/}{https://www.astropy.org/}} package to perform the beam convolution on the simulated line intensity maps. The beam convolution does not change the power spectrum at large scales but reduces the power significantly at small scales, above the scales that correspond to the beam size. 

The shape and amplitude of power spectra of different lines at various redshifts provide valuable information about the properties of the intergalactic medium and galaxy populations. In Figure \ref{fig:simulated_maps}, we present the simulated intensity maps of CII\,158, CO\,(7-6), and OIII\,88 line emissions from the halos at several redshifts corresponding to FYST's EoR-Spec central observational frequencies. We project the intensity grid of the length of $\approx1.3$ $c$Mpc/$h$ for CII\,158, CO\,(7-6), and OIII\,88 lines. We show the three-dimensional power spectra of intensity maps without performing the beam convolution to show both clustering and shot noise terms. However, for the visualization, we convolve the intensity maps with the Gaussian beam, and the FWHM values are varied according to the EoR-Spec on FYST $\nu_{\rm obs}$ \citep{CCAT-prime2021}. For all intensity maps, we consider $M_{\rm min}=10^{10}\,M_\odot$/$h$ at the redshifts mentioned in Figure \ref{fig:HOD_diff}, except the OIII\,88 intensity map at $z\sim 14.5$. Since there are no high-mass halos $\gtrsim 10^{10}\, M_\odot$/$h$ present at such a high redshift, we show the halos whose masses are larger than $10^{9}\, M_\odot$/$h$ for that case. 

At $\nu_{\rm obs}\sim 220\,$GHz, the power spectrum of CO (7-6) lines is $\sim 350$ times larger than CII\,158 power spectrum at $k \sim 0.1$ \,$h$/Mpc, but the ratio becomes 40 to 60 times higher at the shot noise dominated scales, $k\gtrsim 1$  $h$/Mpc. At this frequency corresponding to $z \sim 14.5$, the OIII\,88 signal is negligible as there are very few line-emitting sources. This comparison shows it is impossible to detect OIII\,88 from such high redshift by the ongoing and planned MLIM experiments. However, at $\nu_{\rm obs}\sim 410\,$GHz,  the CII\,158 signal becomes larger than CO (7-6) by a factor of $\sim 6$ at $k= 0.1$ $h$/$c$Mpc and $\sim 2$ at $k= 1$  $h$/Mpc. Furthermore, for the same redshift at $z \sim 7.4$, the OIII\,88 power spectrum is approximately 4.5 and 1.7 times larger than the CII\,158 power spectrum at $k\sim 0.1$ \,$h$/Mpc and 1\,$h$/Mpc, respectively. Therefore, by using two frequency bands, 220 and 280\, GHz, we could detect OIII\,88 and CII\, lines and perform cross-correlation studies as they come from the same sources.

\section{Detectability} \label{sec:detectability}
In this section, we forecast the detectability of CII\,158 and OIII\,88 by considering the specifications of an EoR-Spec experiment. The signal-to-noise ratio is proportional to the number of observed modes present in the survey volume. To determine the number of modes between the wave number $k$ and $k + \Delta k$, we use the following equation:  
\begin{equation}
N_m(k_i,z)= k_i^2 \Delta k_i V_{\rm surv}/ 2\pi^2 \,,
\label{eq:Nm}
\end{equation}
where $k_i$ is the central wave number in the bin width, $\Delta k_i$. The survey volume of an experiment is given by \citep{Gong2017, Dumitru2018}
\begin{multline}
V_{\rm surv}= 3.7 \times 10^{7} (\mathrm{cMpc}/h)^2 \left(\frac{\lambda_{\rm line}}{157.8\,\mu m}\right) \left(\frac{1+z}{8}\right)^{\frac{1}{2}} \\
\times \left(\frac{S_A}{16\, 
\rm {deg^2}} \right) \left(\frac{B_\nu}{20\, \mathrm{GHz}} \right)\,.
\label{eq:Vsurv}
\end{multline}

\begin{figure}
\centering
\includegraphics[width=0.49\textwidth]{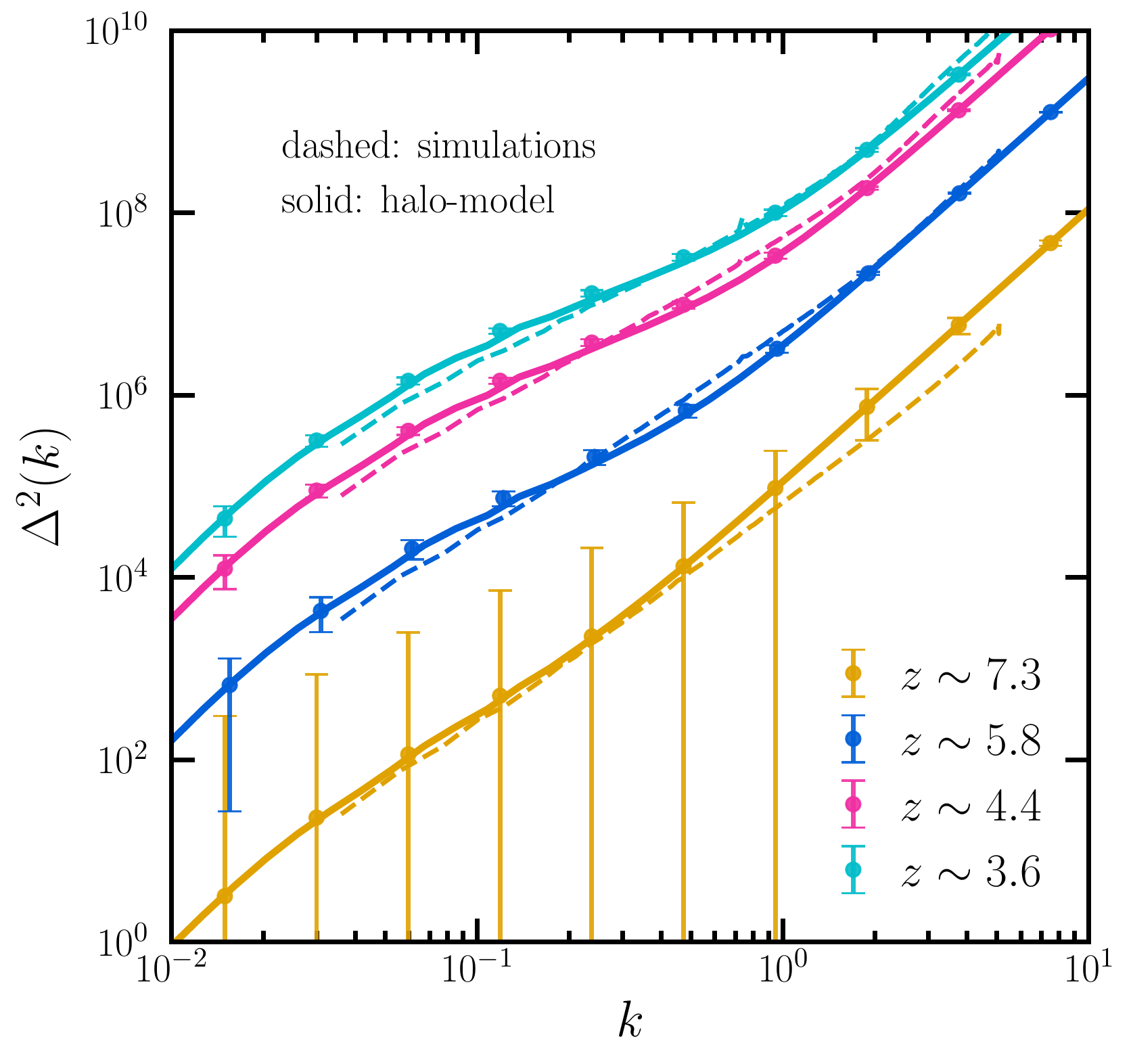} 
\caption{We compare the power spectra of CII\,158 lines at four different redshifts between the halo model and simulations. The figure also shows the error bars for CII\,158 lines forecasted based on the halo model approach for the different frequency bands of the EoR-Spec on the FYST experiment.}
\label{fig:summary}
\end{figure}

In the above equation, $\lambda_{\rm line}$ symbolizes the rest frame wavelength of the line emission, $S_A$ is the effective survey area of an experiment, and $B_{\nu}$ is the frequency bandwidth. For an EoR-Spec-like experiment on FYST, we consider $B_{\nu} = 40$\, GHz for all frequency channels. 

To calculate the covariance matrix for the detection of $P_{\rm line}(k)$, we employ the following formula:
\begin{equation}
\sigma_{\rm line} (z) = \frac{[P^{\rm line} (k, z) + P^{\rm N} (k, z)]^2}{N_{m}(k,z)}\,.
\label{eq:covp3d}
\end{equation}
Here, $P_N$ represents the noise power spectrum. The source of noise can be a combination of white noise, atmospheric noise, and interloper contribution. However, in this paper, we only take the white noise contribution into account to forecast the signal-to-noise ratio (SNR) for the EoR-Spec experiment on FYST \citep{CCAT-prime2021}.

The signal-to-noise ratio, ($S/N$), is given by:
\begin{equation}
{(S/N)}^{2}_{\rm cum}(z) = \sum_i P^2_{\rm line}(k_i, z)/ \sigma^{2}_{\rm line} (z)
\label{eq:cl_sims}
\end{equation}

By employing these equations, we can effectively forecast the detectability of the CII\,158 and OIII\,88 spectral lines, taking into account the white noise as the sole source of noise in our analysis. This allows us to estimate the signal-to-noise ratio for the EoR-Spec experiment on FYST and assess the overall feasibility of detecting these spectral lines.

Figure \ref{fig:summary} presents the detectability status of the CII\,158 power spectrum at different redshifts corresponding to the frequency coverage of the EoR-Spec experiment on FYST. The plot compares the power spectra of CII\,158 between the halo model and simulation, demonstrating the potential of EoR-Spec to detect the CII\,158 signal at different redshifts. The halo mass function used to generate the halo catalogue is the Sheth-Torman mass function, while we use the Tinker mass function for the Halo model approach. The figure also displays the error bars for CII\,158 lines forecasted based on the halo model approach for the different frequency bands of the EoR-Spec. The results of our analysis indicate that the EoR-Spec-like experiment will be able to detect the CII\,158 signal at more than 350 $\sigma$ for ten bins in the range of $k_{\rm min} \sim 0.1$ $c$Mpc/$h$ to $k_{\rm max} \sim 10$ $c$Mpc/$h$ at $z\sim 5.8$. At higher redshifts, the signal-to-noise ratio is expected to be lower, with values of 26, 373, and 295 at $z \sim 7.4, 4.6,$ and $3.4$, respectively. 

Table \ref{tab:snr} summarizes the signal-to-noise ratio (SNR) for the detection of CII,158 at various redshifts corresponding to the FYST's EoR-Spec frequency coverage. Given the considerable uncertainty in the amplitude of the power spectrum, we present SNR forecasts for three different scenarios: optimistic, moderate, and pessimistic, representing the largest, median, and weakest expected signals, respectively. The results demonstrate that an EoR-Spec FYST-like experiment has the potential to detect the CII\,158 signal with high significance, offering a valuable opportunity to constrain theoretical models of galaxy formation and evolution. By probing the complex interplay between various astrophysical processes, these observations could provide crucial insights into the underlying physical mechanisms driving the growth and evolution of galaxies.

\begin{table}[tb]
\centering
\begin{tabular}{l c c c  } 
\hline \hline
 &  \multicolumn{3}{c}{signal-to-noise ratio} \\
 \cline{2-4}
$z_{\rm line}$ & optimistic & moderate & pessimistic  \\
\hline
 7.6  &   284 &  67  &     4\\
 5.8 &   1890  &  460  &  63   \\ 
4.4 &  2041   &  524  &   70   \\
3.6 &   770 &   212  &  28     \\ 
\hline
\end{tabular}
\caption{We quote the forecasted cumulative signal-to-noise ratio of CII\,158 lines at several redshifts for an FYST-like MLIM experiment with taking into account the effect of foregrounds.}
\end{table}\label{tab:snr}

\section{Conclusion} \label{sec:conclusion}
In the study of galaxy formation and line intensity mapping, uncertainties in astrophysical modelling of line intensity signals and variations in star formation histories within dark matter halos are crucial topics. In our research, we have developed a package that brings together various models to enable comparison and facilitate the elimination of models when making MLIM observations. The $\limpy$ package is a semi-numerical code that allows for the modelling of CII\,158, OIII\,88, and different CO J-ladder transitions in a single framework. We have implemented several star formation models, including those inferred from analytic prescriptions and state-of-the-art simulations such as $\tng$ and $\universemachine$, as well as empirical relations from the abundance matching approach. We assume that the SFR serves as a proxy for line luminosities and have included various SFR$-L_{\rm line}$ scaling relations based on several best-fit models. Our primary objective in this study is to provide a tool for investigating the astrophysical and cosmological information derived from line intensity mapping while taking into account the modelling uncertainties inherent in such an approach. By integrating various models and scaling relations in a single framework, our approach can help interpret the observed MLIM signal. 

The $\limpy$ package not only allows for the modelling of line intensity maps but also enables the exploration of cosmological parameters and their effects on these maps. Halo catalogues generated from various simulations, such as N-body or cosmological hydro simulations, can be inputted into the code to investigate the impact of these parameters on the line emissions. The code efficiently paints halos with line emissions based on the SFR and line luminosity models, making it an ideal tool for performing MCMC analyses on simulations to be constrained by MLIM observations. Furthermore, the simulated maps produced by $\limpy$ can be used to determine optimal statistics for analyzing observed MLIM data. For instance, future studies may employ voxel intensity distribution (VID) to analyze the simulated multi-line intensity maps outputted by the package \citep{COMAP:2018kem, Breysse:2022alx, Sato-Polito:2020cil}. Overall, $\limpy$ provides a versatile and powerful tool for studying both astrophysical and cosmological parameters at the map level and can facilitate current and future MLIM observations. 

The CII\,158 line emission, which exhibits a large scatter, raises critical questions about interpreting the MLIM signal at these redshifts during observations. The uncertainties in the astrophysical parameters pose a challenge in accurately constraining the properties of the intergalactic medium and galaxy populations responsible for the observed signals. Thus, low-noise MLIM observations are crucial to obtaining a robust analysis of the data and constraining the astrophysical parameters. This highlights the need for improving the modelling of the CII\,158 power spectrum and other line intensities to maximize the scientific returns of MLIM experiments.

The high signal-to-noise ratio attainable through an EoR-Spec-like experiment for the CII\,158 signal makes it an ideal tool for probing the parameters associated with the reionization process, such as the ionized bubble size and mean free path of photons. These observations could be used to reconstruct the luminosity function of galaxies and provide insight into the ionizing sources. Moreover, the frequency overlap of the EoR-Spec allows for the inter-line cross-correlation of CII\,158 (220 GHz) with OIII\,88 (410 GHz) to obtain a snapshot of the Universe at $z \sim 7.4$. However, this analysis is subject to potential issues such as interloper contamination and the effect of beam convolution, which we neglected in our forecasts. Consequently, while our forecasts for detecting CII\,158 are optimistic, further work is required to account for these factors and obtain a more accurate estimation of the detectability of the CII\,158 signal.

However, the rotational emissions from CO molecules at low redshifts dominate over the CII\,158 emissions from high redshifts, presenting a significant obstacle to studying the reionization epoch using CII\,158. Additionally, extragalactic foregrounds, such as broadband emission from the cosmic infrared background (CIB), contribute to the contamination. To minimize contamination, linear combinations of maps reconstructed using different channels can be used. This approach can significantly reduce the bias for CII\,158 detections. However, to obtain more realistic predictions, future work will explore bias due to interlopers, instrumental and atmospheric noise, and will build estimators to estimate the signal in their presence.

\section{Acknowledgement}
AR would like to thank Anthony Challinor, Steve Choi, Andrea Lapi, Dominik Reichers, and Gordon Stacey for helpful discussions. AR is partially supported by the CCAT-prime collaboration. DV acknowledges the REU program through the CCAPS at Cornell University under the NSF award NST/AST-1950324. NB acknowledges support from NSF grant AST-1910021 and NASA grants 21-ADAP21-0114 and 21-ATP21-0129. AvE acknowledges support from NASA grants 22-ADAP22-0149 and 22-ADAP22-0150.

% References %
\bibliographystyle{aasjournal}
\bibliography{citation}

\newpage

\appendix

\section{Fitting coefficients in a  multi-line power spectrum analysis} \label{sec:fitting_coeff}
In this section, we present a parameterization of the power spectra of various spectral lines that relies solely on the matter power spectrum at different redshifts. The modelling of SFR and $L_{\rm line}-SFR$ has a large degree of uncertainty across a wide range of redshifts. Therefore, we use a simple two-parameter fit to model the power spectra of multi-lines as
\begin{equation}
\frac{P^{\rm fit}_{\rm line}(k, z)}{(Jy/Sr)^2 (h^{-3}{Mpc}^3)}=  10^{A(z)}\times \frac{P_{m} (k,z )}{h^{-3}Mpc^3} + 10^{B(z)} \,. \label{eq:fitting_func}
\end{equation}

Here, $P_{\rm line}^{\rm fit}$ refers to the power spectra of a particular line emission. $A(z)$ and $B(z)$ are two free parameters that we fit with the power spectra that we calculate from simulations. The unit of $P_{\rm m}(k,z)$ is $h^{-3}Mpc^3$. The parameter $A(z)$ scales the clustering term, while $B(z)$ is analogous to the noise term of power spectra. By fitting these parameters, we can construct a model that describes the power spectra of different line emissions at different redshifts that will be useful to estimate the multi-line intensity power spectrum quickly. 

We utilize the large-scale structure information from the $\tng$ simulation to paint multiple lines onto the halo catalogues using the $\limpy$ package. We use the TNG300 simulation, in which the box size is $\sim 205$ $c$Mpc/$h$, and we set the minimum mass of the halos, $M_{\rm halo} \gtrsim 10^{10}\,M_\odot$/$h$. We then paint the halos with CII\,158, OIII\,88, and full J-ladder  transitions from CO\,(1-0) to CO\,(13-12) lines using the $\limpy$ package. By painting the halos with these lines, we can better understand their scaling relation with the matter power spectrum and how they evolve over cosmic time. To fit the power spectrum to the simulation results, we utilize the fitting parameters $A(z)$ and $B(z)$, which are provided in Table \ref{tab:fitting_params} at redshifts from 0 to 10. This fitting function presented in Equation \ref{eq:fitting_func} could be useful in constraining cosmological parameters that affect the matter power spectrum while accounting for the complex astrophysics captured by the free parameters $A(z)$ and $B(z)$. This approach enables us to model the power spectra of various lines without relying on uncertain models of star formation and the $L_{\rm line}-SFR$ relation across a broad range of redshifts.

\begin{table*}[h]
\centering
\footnotesize
\setlength{\extrarowheight}{6pt}
\begin{tabular}{|c|c|c|} 
\hline
 &  $A(z)$ & $B(z)$\\ 
line name & $z =$ [0, 1, 2, 3, 4, 5, 6, 7, 8, 9, 10]  & $z =$[0, 1, 2, 3, 4, 5, 6, 7, 8, 9, 10] \\\hline

CII\,158 & [4.51, 6.81, 7.25, 7.12, 6.76, 6.34, 5.80, 5.22, 4.52, 3.43, 2.32] &  [6.56, 8.47, 8.75, 8.48, 7.97, 7.41, 6.81, 6.25, 5.69, 4.87, 3.98]  \\

OIII\,88 & [5.10, 7.37, 7.86, 7.79, 7.51, 7.17, 6.74, 6.33, 5.86, 4.94, 3.96]   &  [7.01, 8.90, 9.22, 9.01, 8.55, 8.03, 7.47, 6.95, 6.46, 5.72, 4.94]  \\ 

CO\,(1-0) & [1.95, 3.73, 3.97, 3.80, 3.52, 3.23, 2.86, 2.51, 2.05, 1.16, 0.20]  &  [3.83, 5.23, 5.31, 4.97, 4.48, 3.98, 3.47, 3.02, 2.58, 1.85, 1.08] \\ 

CO\,(2-1) & [2.67, 4.74, 5.08, 4.94, 4.61, 4.26, 3.85, 3.46, 2.99, 2.06, 1.08]  &  [4.64, 6.33, 6.52, 6.22, 5.71, 5.18, 4.62, 4.12, 3.63, 2.88, 2.07] \\ 

CO\,(3-2) &  [3.26, 5.19, 5.48, 5.33, 5.02, 4.70, 4.31, 3.94, 3.47, 2.56, 1.59]  &  [5.19, 6.74, 6.88, 6.56, 6.06, 5.55, 5.01, 4.53, 4.06, 3.32, 2.53] \\

CO\,(4-3) &  [3.08, 5.18, 5.54, 5.40, 5.07, 4.72, 4.29, 3.89, 3.42, 2.49, 1.51]  &  [5.05, 6.78, 6.99, 6.70, 6.19, 5.65, 5.08, 4.57, 4.09, 3.32, 2.51] \\

CO\,(5-4) & [2.91, 5.11, 5.50, 5.37, 5.03, 4.66, 4.21, 3.80, 3.33, 2.38, 1.39] & [4.91, 6.73, 6.98, 6.69, 6.18, 5.64, 5.05, 4.53, 4.03, 3.25, 2.43]  \\

CO\,(6-5) & [2.60, 4.82, 5.23, 5.09, 4.75, 4.38, 3.93, 3.51, 3.04, 2.09, 1.10]   &  [4.60, 6.45, 6.71, 6.43, 5.91, 5.37, 4.78, 4.25, 3.75, 2.97, 2.15] \\

CO\,(7-6) &  [2.32, 4.70, 5.16, 5.04, 4.68, 4.28, 3.80, 3.37, 2.87, 1.91, 0.91]  &  [4.36, 6.35, 6.67, 6.41, 5.9 , 5.33, 4.72, 4.17, 3.65, 2.85, 2.01] \\

CO\,(8-7) & [2.08, 4.40, 4.84, 4.72, 4.36, 3.97, 3.50, 3.07, 2.59, 1.63, 0.63] & [4.10, 6.04, 6.34, 6.08, 5.56, 5.00, 4.40, 3.85, 3.34, 2.55, 1.72]  \\

CO\,(9-8) & [2.09, 4.33, 4.75, 4.61, 4.27, 3.89, 3.44, 3.02, 2.54, 1.59, 0.60] & [4.09, 5.96, 6.23, 5.95, 5.44, 4.89, 4.3 , 3.76, 3.26, 2.48, 1.65] \\

CO\,(10-9) &  [1.94, 4.23, 4.67, 4.54, 4.19, 3.80, 3.34, 2.91, 2.43, 1.47, 0.48]  &  [3.95, 5.87, 6.16, 5.89, 5.38, 4.82, 4.22, 3.68, 3.17, 2.38, 1.55] \\

CO\,(11-10) & [1.69, 3.86, 4.25, 4.11, 3.77, 3.40, 2.97, 2.56, 2.08, 1.14, 0.16] & [3.68, 5.47, 5.72, 5.43, 4.92, 4.37, 3.79, 3.27, 2.78, 2.01, 1.19] \\

CO\,(12-11) & [1.50, 3.85, 4.30, 4.18, 3.82, 3.43, 2.96, 2.52, 2.03, 1.07, 0.08] & [3.53, 5.50, 5.81, 5.55, 5.03, 4.47, 3.86, 3.31, 2.80, 2.01, 1.17]  \\

CO\,(13-12) & [1.67, 3.72, 4.05, 3.90, 3.58, 3.24, 2.83, 2.44, 1.97, 1.05, 0.07] & [3.63, 5.30, 5.49, 5.18, 4.68, 4.15, 3.59, 3.09, 2.61, 1.85, 1.05]  \\ \hline
\end{tabular}
\caption{We quote the values of parameters $A(z)$ and $B(z)$ for various line emissions fitted to the power spectra obtained from simulations spanning across the redshift range from 0 to 10.}
\label{tab:fitting_params}
\end{table*}

\section{Multi-line power spectrum} 
\label{sec:app_power}
\begin{figure*}[h]
\centering
\includegraphics[width=0.9\textwidth]{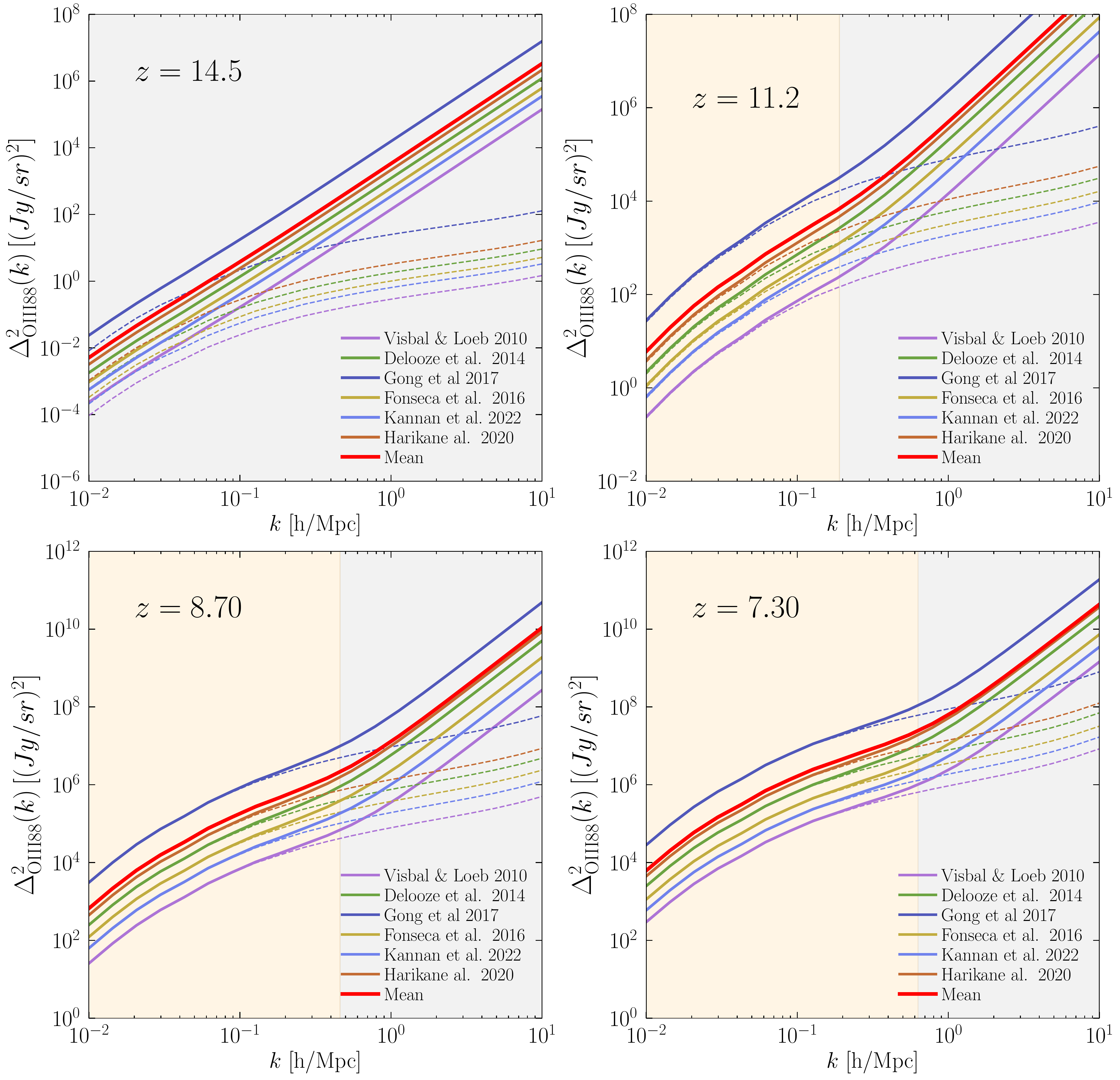}
\caption{The power spectra of OIII\,88 lines at redshifts 14.5, 11.2, 8.7, and 7.3. Dashed lines show the signal from the clustering term only, and solid lines describe the total signal comprising both the clustering and shot noise terms. The solid red lines at each panel are the best-fit (mean) line of all these models.}
\label{fig:app:ps_OIII}
\end{figure*}

\begin{figure*}[h]
\centering
\includegraphics[width=\textwidth]{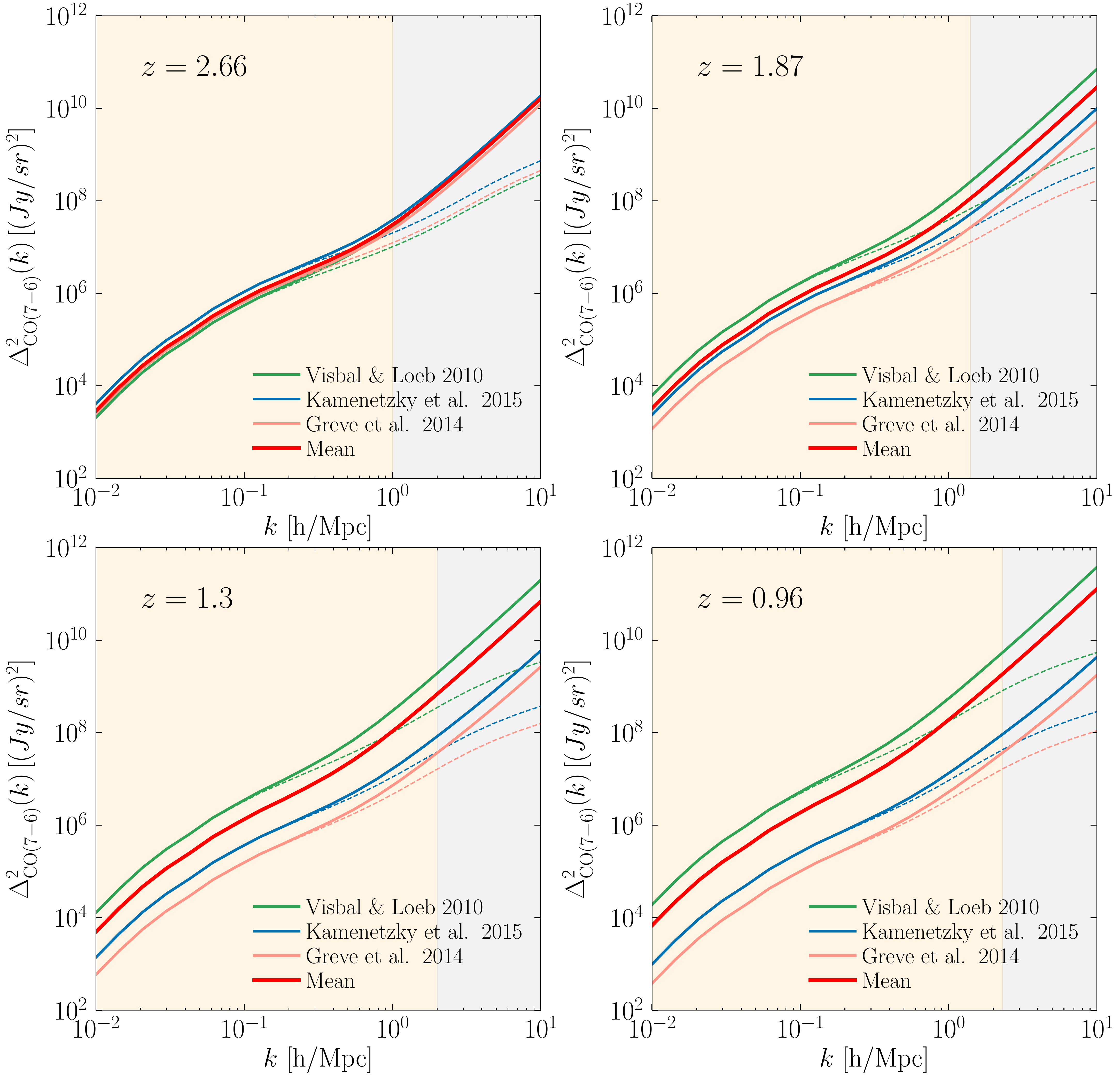} 
\caption{We show the power spectra of CO\,(7-6) lines obtained using various models. The \texttt{Silva15} star formation rate (SFR) model is used to determine the signal at all redshifts to maintain consistency in the analysis.}
\label{fig:app:ps_CO}
\end{figure*}

We determine the power spectra of OIII\,88 and CO\,(7-6) at the redshift corresponding to the four frequency channels of FYST's EoR-Spec. By studying the power spectra of OIII\,88 and CO\,(7-6) at these redshifts, we aim to understand the properties of these lines and their role in galaxy formation and evolution. Estimation of power spectra of other lines is important to understand the level of contamination of a particular line of interest. To demonstrate our findings, we present the power spectra of OIII\,88 and CO\,(7-6) in Figure \ref{fig:app:ps_OIII} and \ref{fig:app:ps_CO}, respectively. These figures showcase the power spectra obtained from our analysis and highlight the potential insights that can be gained from a multi-line approach to studying galaxy evolution.

\end{document}